**Ride-pooling potential under alternative spatial demand patterns**

**Jaime Soza-Parra, Corresponding author**
Institute of Transtportation Studies
University of California, Davis
1605 Tilia Street, Davis, CA 95616, United States of America
Email: jsozaparra@ucdavis.edu

**Rafał Kucharski**
Transportation & Planning Department
Delft University of Technology
P.O. Box 5048
2600 GA Delft, The Netherlands
Email: r.m.kucharski@tudelft.nl

**Oded Cats**
Transportation & Planning Department
Delft University of Technology
P.O. Box 5048
2600 GA Delft, The Netherlands
Email: o.cats@tudelft.nl

**Declaration of interests:** None

**Acknowledgements**

This research was supported by the CriticalMaaS project (grant number 804469), which is financed by the European Research Council and Amsterdam Institute for Advanced Metropolitan Solutions.

## Ride-pooling potential under alternative spatial demand patterns

### Abstract

Shared rides are often considered to be a promising travel alternative that could efficiently pool people together while offering a door-to-door service. Notwithstanding, even though demand distribution patterns are expected to greatly affect the potential for ride-pooling, their impact remains unknown. In this study we explore the shareability of various demand patterns. We devise a set of experiments tailored to identify the most promising demand patterns for introducing ride-pooling services by varying the number of centers, the dispersion of destinations around each of these centers and the trip length distribution. When matching trips into rides, we do not only ensure their mutual compatibility in time and space but also that shared rides are only composed by travellers who find the ride-pooling offer to be more attractive than the private ride-hailing alternative given the trade-offs between travel time, fare and discomfort. We measure the shareability potential using a series of metrics related to the extent to which passenger demand can be assigned to shared rides. Our findings indicate that introducing a ride-pooling service can reduce vehicle-hours by 18-59% under a fixed demand level and depending on the concentration of travel destinations around the center and the trip length distribution. System efficiency correlates positively with the former and negatively with the latter. A shift from a monocentric to a polycentric demand pattern is found to have a limited impact on the prospects of shared rides.

### 1. Introduction

Ride-hailing services have become part of the urban mobility landscape across the world. There is a fervent debate on the impacts of ride-hailing services, especially in relation to their competition with public transport services (Boisjoly et al. 2018) and contribution to traffic congestion (Erhardt et al. 2019). Shared rides are likely, however, to be more consistent with policy goals such as improved accessibility and affordability as well as reducing congestion and related externalities. For passengers to choose for ride-pooling over a private ride, a discount is offered in order to compensate for the potentially induced discomfort and delay. From platform and drivers' perspective, ride-pooling may increase the overall market share of ride-hailing and thus generate additional income. Notably, even though ride-hailing operations facilitate the emergence of ride-pooling services, their market shares remain insofar low (Li et al. 2019a, Young et al. 2020). Moreover, it is estimated that only about half of the trips which took place using the pooled service were actually shared in the case of Toronto (Young et al. 2020). It is thus pertinent to identify under what circumstances are ride-pooling most likely to attain a significant market share and result with efficiency gains. In this study we set to identify the most promsing spatial demand patterns for introducing ride-pooling services, not only by resulting with travel requests that can be shared, but that will ultimately will also be chosen by travellers over alternative private rides.

For ride-hailing companies to offer an attractive ride-sharing service it is crucial to ensure that there is a high likelihood that trips can be pooled together into shared rides. The latter depends on the availability of travel requests that are mutually compatible in terms of their spatial and temporal constraints so that they do not impose prohibitive detours or delays for any of the co-riders. As can be expected, Tachet et al. (2017) show that the higher the demand volume, the higher the likelihood that trips can be matched.

Moreover, using taxi data from various cities around the world, Tachet et al. found that there is a common pattern where an increase in demand results with a rapid increase in the ability to bundle trips which is quickly saturated. Taxi demand accounts however for a small share of urban mobility and may exhibit unique characteristics which are not representative of the overall demand pattern. This has given rise to the notion of critical mass that needs to be obtained for ride-pooling to become attractive. Some past studies have pointed to the importance of travel demand directionality in explaining fleet utilization (Narayan et al. 2021). Notwithstanding, even though demand distribution patterns are expected to greatly affect the potential for ride-pooling, their impact remains unknown.

A stream of empirical studies have investigated the relations between urban and road network structure, land-use distribution, travel demand and transport performance. Ewing and Cervero (2010) performed a meta-analysis of studies analyzing behavior at the individual level from across the United States. Several studies conducted a regression analysis at the urban agglomeration level, based on car travel (Ewing et al. 2018) or mobile phone location (Bassolas et al. 2019) data from US and location-based navigation service data for Chinese (Li et al. 2019b) cities, and public transport ridership data from across Europe (Blafoss Ingvardson and Anker Nielsen 2018). Hierarchical agglomerative clustering was applied to identify the relation between urban characteristics and mobility patterns for different clusters of global cities (Oke et al. 2019). The results of these studies offer conflicting conclusions as to whether compact (as opposed to dispersed) city centers and the related travel demand patterns are associated with increased or decreased travel mileage and traffic congestion.

Analytical and simulation studies have also been employed for establishing the relation between the underlying demand pattern and the performance of various transport modes. Analytical models estimated the macroscopic flow diagram properties for a concentric city model (Tsekeris and Geroliminis 2013) or identified the public transport network structure for various urban structure characteristics such as trip dispersion and the importance of the urban center and sub-centers (Fielbaum et al. 2016) as well as for the levels of compactness for a ring-radial city (Badia 2020). Network evolution models have demonstrated that structurally distinguished demand patterns facilitate the emergence of distinctive public transport networks for a ring-radial agglomeration (Cats et al. 2020). These studies offer a variety of modelling approaches and insights on the relation between demand patterns, travel demand and service performance. However, there is lack of knowledge as to the ramifications of those for the prospects of ride-pooling. In a related study, Wang and Zhang (2020) applied an agent-based shared automated vehicle simulator for various US cities and regressed the results in relation to urban agglomeration characteristics. They found that job density and land-use diversity contribute to the percentage of pooled trips. The simulation model used randomly assigns travellers with a dummy variable indicating whether they are willing to share or not based on an input rate parameter and pools two travellers if they are both assumed willing. However, in reality passengers are expected to choose between private rides, shared rides and alternative modes based on the discount offered for ride-pooling and its trade-off with the delay and discomfort induced.

The key research question of this study is: how ‘shareable’ are various spatial demand patterns? To this end, we devise a set of experiments tailored to identify the most promising demand patterns for introducing ride-pooling services. To disentangle the critical mass effect needed to induce pooling from the spatial pattern significance we keep the demand levels fixed through the experiments. We generate a broad range of synthetic demand patterns by varying the number of centers, the dispersion of destinations around each of these centers and the trip length distribution. Subsequently, we identify feasible matches between travellers and identify feasible pooled rides where two or more travellers share the same vehicle. We then apply an exact matching algorithm (Kucharski and Cats, 2020) and examine the consequences of alternative demand patterns on travellers' choice for ride-pooling, its level-of-service and system performance.

The remaining of this paper is organized as follows. In the next section we detail the process of generating alternative demand patterns, the algorithms used for matching travel requests into rides and the indicators proposed for quantifying the shareability potential of a given demand pattern. Next, we describe the experimental set-up based on the Amsterdam case study and compare the results obtained for different scenarios as well as their spatial variations and distributional effects. We conclude with a discussion of the key findings and their implications as well as suggestions for further research.

## 2. Methodology

In this section, we describe the sequence of steps that we undertake to generate and assess our set of demand distribution scenarios. First, we describe how we generated a parametric set of demand patterns (Section 2.1). Second, we outline how we calculate the shareability potential of each of those scenarios, followed by the set of key performance indicators used for assessing the shareability potential ingrained in each of the demand patterns (2.2).

### 2.1 Synthetic Demand Generation

We generate a set of synthetic scenarios that are meant to reflect different demand patterns that might prevail in urban areas. In the following, we propose characterizing the urban demand distribution - and generating variations thereof by means of determining the following design variables:

(i) number of centers,

(ii) density of destinations around each center,

(iii) trip length distribution.

As illustrated in Figure 1, the demand generation consists of three consecutive stages. We first determine the number of centers, around which we generate destinations with a given density. We then connect generated destinations with predefined origins while controlling for trip lengths. The second and third steps are controlled via parameters of spatial distributions ranging from concentrated to disperse and from short to long trips, respectively. Such synthetic demand patterns cover urban areas consisting of one or more centers; with both highly concentrated and uniformly dispersed destinations, and; with low impedance (travellers likely to travel long distances) and high impedance (where travellers find destinations nearby). This process is detailed in the following.

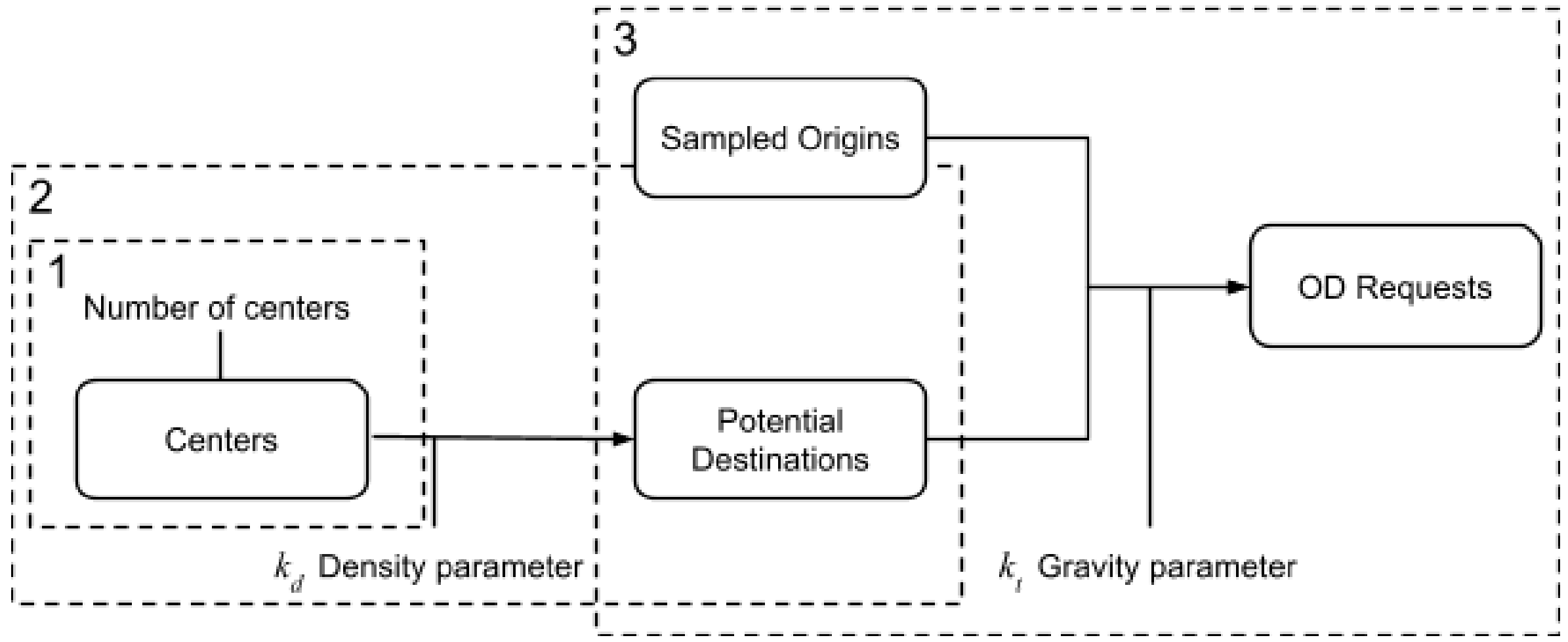

**Figure 1. Synthetic demand generation approach**

First, we determine centers in a predefined set of locations. The centers are then fixed in order to obtain unbiased statistics such as the average distance. Furthermore, the centers are categorized in terms of relative importance, so that the same set of centers is considered in each scenario which involves a given number of centers.

Second, for each center, potential destinations $d$ are generated following a Gamma distribution with shape parameter $k_d$ and scale parameter $s_d$, hence its probability density function is as follows:

$$f(d;k_d,s_d)=\frac{d^{k_d-1}\cdot e^{-\frac{d}{s_d}}}{s_d^{k_d}\cdot\Gamma(k_d)} \tag{1}$$

Where $\Gamma(k_d)$ corresponds to the Gamma function evaluated at the shape parameter $k_d$:

$$\Gamma(k_d)=\int_0^{\infty} x^{k_d-1}\cdot e^{-x}\,dx \tag{2}$$

The scale parameter is fixed to represent the city under consideration and the shape parameter reflects the density. This probability density function was selected as it describes well the distance from a bivariate normal distributed coordinate to its center. The Gamma distribution is a member of the natural exponential family of distributions. Hence, when the shape parameter is lower than one, the distribution is exponentially shaped and asymptotic to both axes, when it is equal to one, it corresponds to an exponential distribution with mean equals to the scale of the distribution, and when it is larger than one the distribution is skewed and varies with shape. This means that the lower this parameter is, potential destinations are more concentrated around each of the centers. Four specific distributions are illustrated in Figure 2.

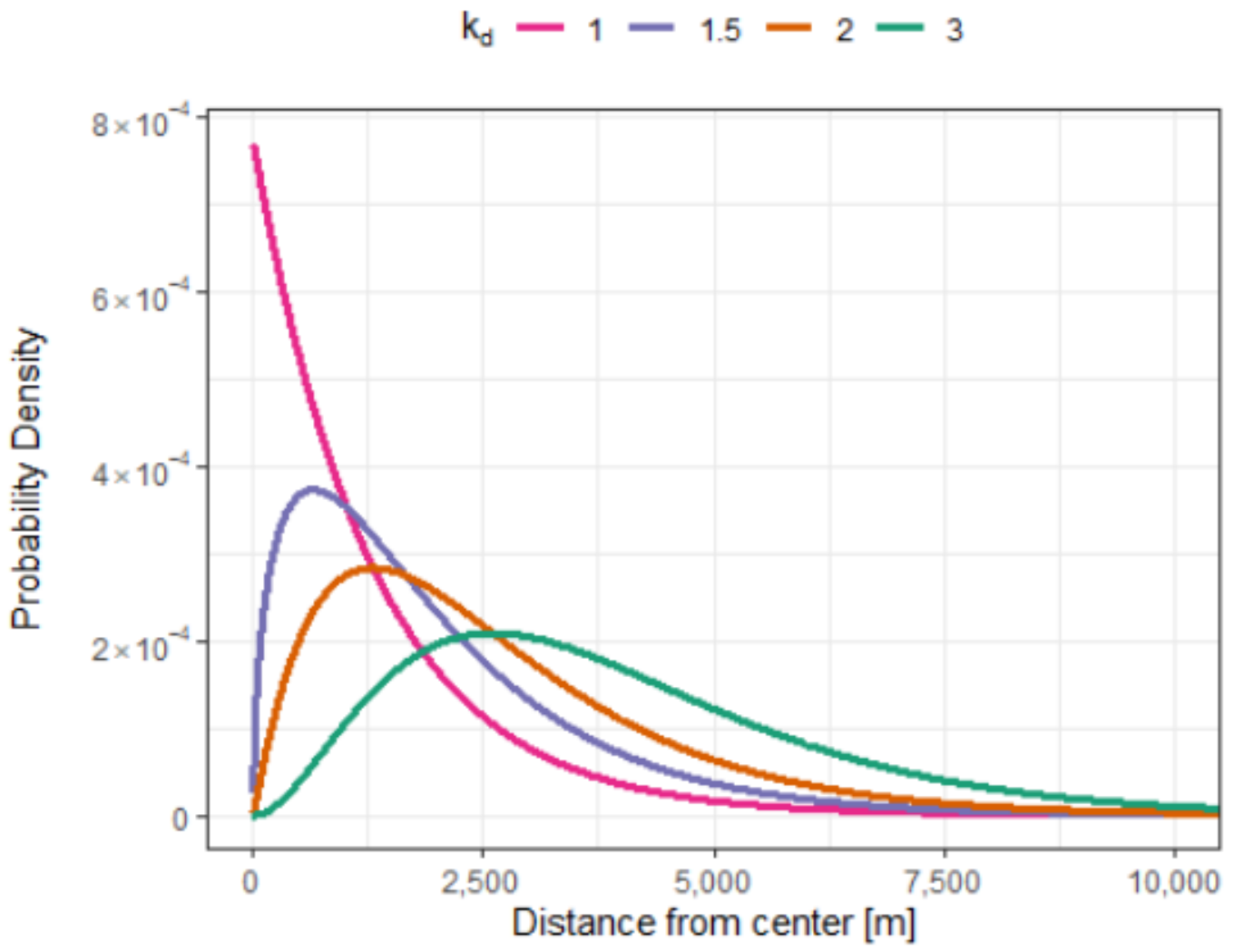

**Figure 2 Destinations' distance from center distribution for four different $k_d$ values**

Third, we match destinations with origins. In this study, we use the actual spatial distribution of trip origins. To create an origin-destination matrix, fixed origins are matched with synthetically generated destinations based on the trip length distribution parameter. Similarly to the second stage, this trip length distribution parameter corresponds to the shape parameter of an independent Gamma probability density function, for which its scale parameter has been fixed. As before, we define the distance of these generated trips, $d$, to follow a Gamma distribution with shape parameter $k_t$ and scale parameter $s_t$. The lower parameter $k_t$ is, the destinations are found closer to their origins. The probability density function is thus defined as follows:

$$f\left(d;k_t,s_t\right)=\frac{d^{k_t-1}\cdot e^{-\frac{d}{s_t}}}{s_t^{k_t}\cdot\Gamma\left(k_t\right)} \quad (3)$$

As before, four different distributions are shown for illustration in Figure 3.

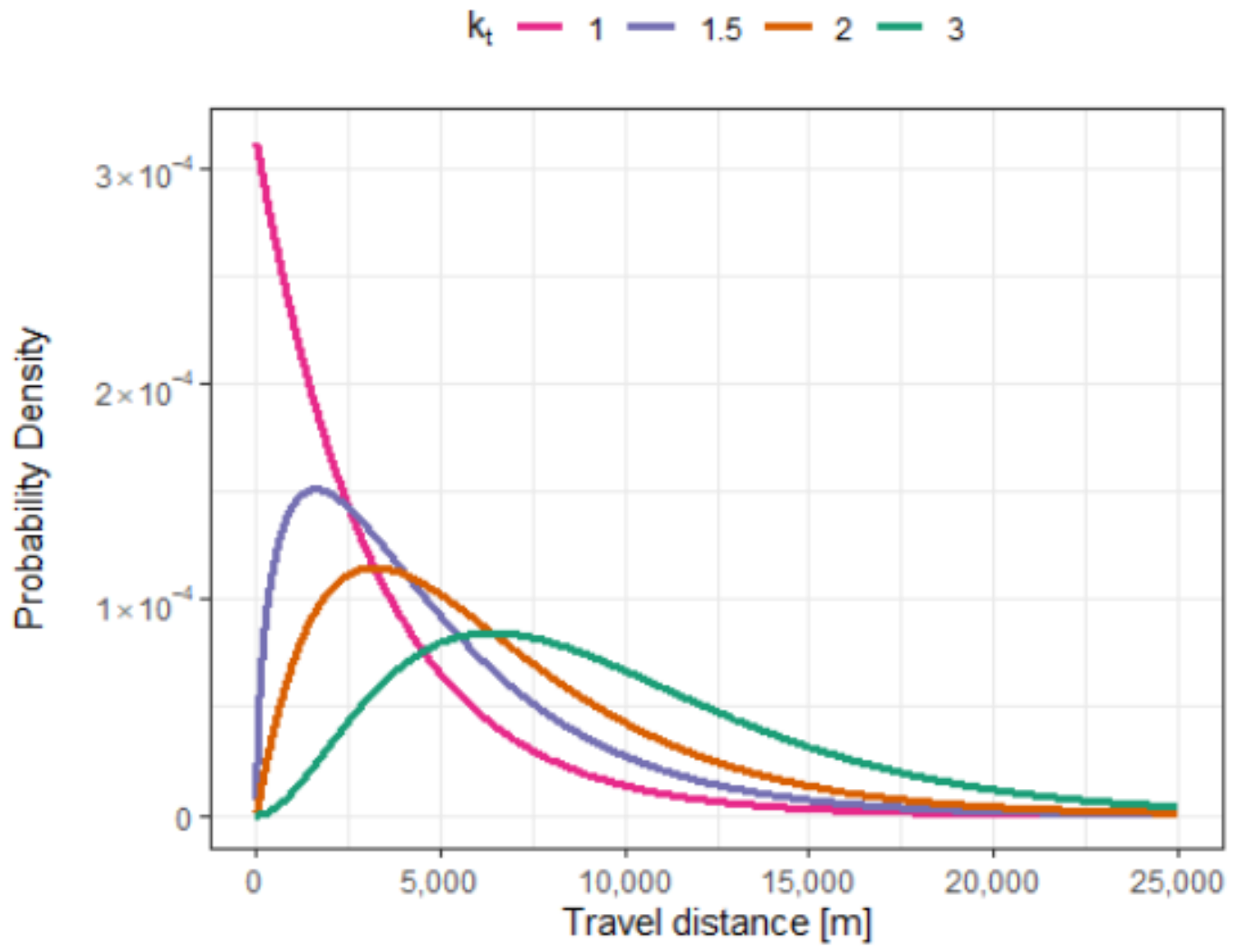

**Figure 3 Synthetic travels' distance distribution for four different $k_t$ values**

Finally, we obtain a synthetic set of requests, where for each predefined origin we generate a destination in such a way that the resulting demand patterns follow the above shown spatial distributions.

### 2.2 Quantifying Shareability potential

For each of the generated synthetic travel demand patterns we are interested in quantifying their shareability potential. To this end, we apply the ExMAS algorithm (Kucharski and Cats 2020) which pools single trips into attractive shared rides. This algorithm is suited for strategic planning purposes, focusing on the demand side, rather than operational platform considerations. It uses as input the demand pattern (set of trip requests with their origin, destination and departure time) and identifies feasible pooled rides for given sets of behavioural parameters (value-of-time, and willingness to share) and design variables (discount offered for pooled rides). Based on this, ExMAS identifies attractive pooled rides, for which the detour and delay are compensated with the reduced fare. Since it is deterministic, it allows computing shareability metrics for relatively large-scale demand patterns using compensatory behavioural decision making rules. Mind that in ExMAS we consider a flat discount for sharing in a distance-based fare system, which makes the shareability results sensitive to trip lengths (for long trips longer detours are accepted due to greater absolute discount for pooling). The ExMAS algorithm yields an optimal matching of travellers to shared-rides based on which the following metrics are computed:

1. Total vehicle hours, $T$ - total time spent with travellers to satisfy the demand (deadheading and other empty trips are not considered). This allows assessing the efficiency of ride-pooling for the supplier (platform operator or service provider), which, in principle, shall be smaller than for non-pooled scenarios.

2. Total passenger cost, $C$ - cost of all travellers, which combines the disutility of time (in-vehicle and waiting) with the monetary terms (fare, possibly reduced due to pooling), which, in principle, shall be reduced thanks to pooling.

3. Occupancy, $O$ - ratio of vehicle hours ($T$) to passenger hours ($T_{pass}$), supports investigating how compactly the travel demand is pooled. We use it as a key indicator for studying the efficiency attained by pooling.

For vehicle hours and passenger costs, we compare the values obtained when allowing for pooling versus the case where a ride-pooling service is not available. We report the relative differences that result from sharing $\Delta C$, $\Delta T$ as follows:

$$\Delta C = \frac{C - C_{base}}{C_{base}} ; \Delta T = \frac{T - T_{base}}{T_{base}}$$

This set of metrics allows for the sound assessment of the shareability potential associated with each of the demand patterns.

## 3. Application

In this section we first describe the experimental set-up devised to address our research question in (Section 3.1). Then we present the results, first in terms of the resulting synthetic demand patterns (3.2.1) followed by aggregate (3.2.2) and distributional (3.2.3) shareability indicators.

### 3.1 Experimental set-up

The experiment is configured around the case of Amsterdam, the Netherlands, where 1,000 trip requests are generated during a one hour period. This aims to replicate ride-hailing operation in the city, as it starts with 1000 origins sampled from the actual Amsterdam demand pattern extracted from a nation-wide synthetic demand model (Arentze and Timmermans, 2004) and keep them fixed across the experiments. As explained in the previous section, the first step consists of defining the number of centers around which the synthetic demand is pivoted. We consider up to four possible centers, namely Dam Square, Station Zuid, Concertgebouw (Museumplein) and Sloterdijk (their locations can be seen later in Figure 7). These four centers constitute major travel attraction areas during the morning peak hour and we therefore choose to use them in our experiments as the anchors for demand attraction. In addition, these four centers are employed in the abovementioned order, meaning that when we simulate a monocentric city, we always consider Dam Square as the sole center of attraction, whereas when considering two centers, we include Dam Square and Station Zuid, and so forth.

In terms of the other two attributes, namely $k_t$ and $k_d$ , we define them around the existing origin-destination pattern in Amsterdam. If we impose that Amsterdam is a monocentric city, meaning destinations are concentrated with the Dam Square as the sole center of gravity, we obtain a trip length distribution parameter, $k_t$ , of approximately 1.5, and a destination density around the center, $k_d$ , of around 2.5. Since in reality, destinations in Amsterdam and elsewhere radiate around a multiplicity of centers, we experiment with increasing the travel time distribution parameter $k_t$ and decreasing the density parameter around centers $k_d$. For completeness, we also consider additional scenarios with a lower $k_t$ and a higher $k_d$ than those observed for Amsterdam in reality.

Consequently, the experiments are run over the following grid of parameter values:

$n$ ={1, 2, 3, 4},

$k_t$ ={1, 1.5, 2, 3} ,

$k_d$ ={1, 1.5, 2, 2.5, 3},

Our experimental design results in a total of 80 (4x4x5) scenarios to be simulated. Since each scenario is associated with a random demand generation (whereas the ExMAS ride-pooling algorithm is deterministic), we replicate scenarios to obtain statistically significant results. Based on the distribution of the output metrics, 10 replications per scenario were found sufficient to ensure a maximum allowable error of less than 5%.

### 3.2 Results

In this section we first report the demand patterns generated for the study and then explore their shareability prospects.

#### 3.2.1 Demand patterns description

Each simulation is specified based on the number of centers and a combination of both shape parameters for the distance between the potential destinations and each center, $k_d$, and the trip length distribution, $k_t$. In the following, we first report the characteristics of the demand patterns generated in order to better understand the underlying causes of the shareability findings and the spatial variations thereof that are reported in subsequent sections.

We first analyse the distance of each potential destination generated in a monocentric city to its center (Figure 4, left), followed by the trip length distribution (Figure 4, right). The boxplots - the lines spanning from the 10th to the 90th percentile and from the 25th to the 75th percentiles define the box limits, the median is represented by the middle line and the dot represents the average value - illustrate, as expected, that destinations are more dispersed when increasing $k_t$, not only the mean distance from the center, but also its variability, increasing (Figure 4, left). In terms of the distribution of potential destinations, we see that the average distance to centroid increases with $k_d$, especially when $k_t$ = 3. In terms of average trip distance, we see that it increases significantly when increasing $k_t$, as expected, and at a lower rate when increasing $k_d$ (Figure 4, right).

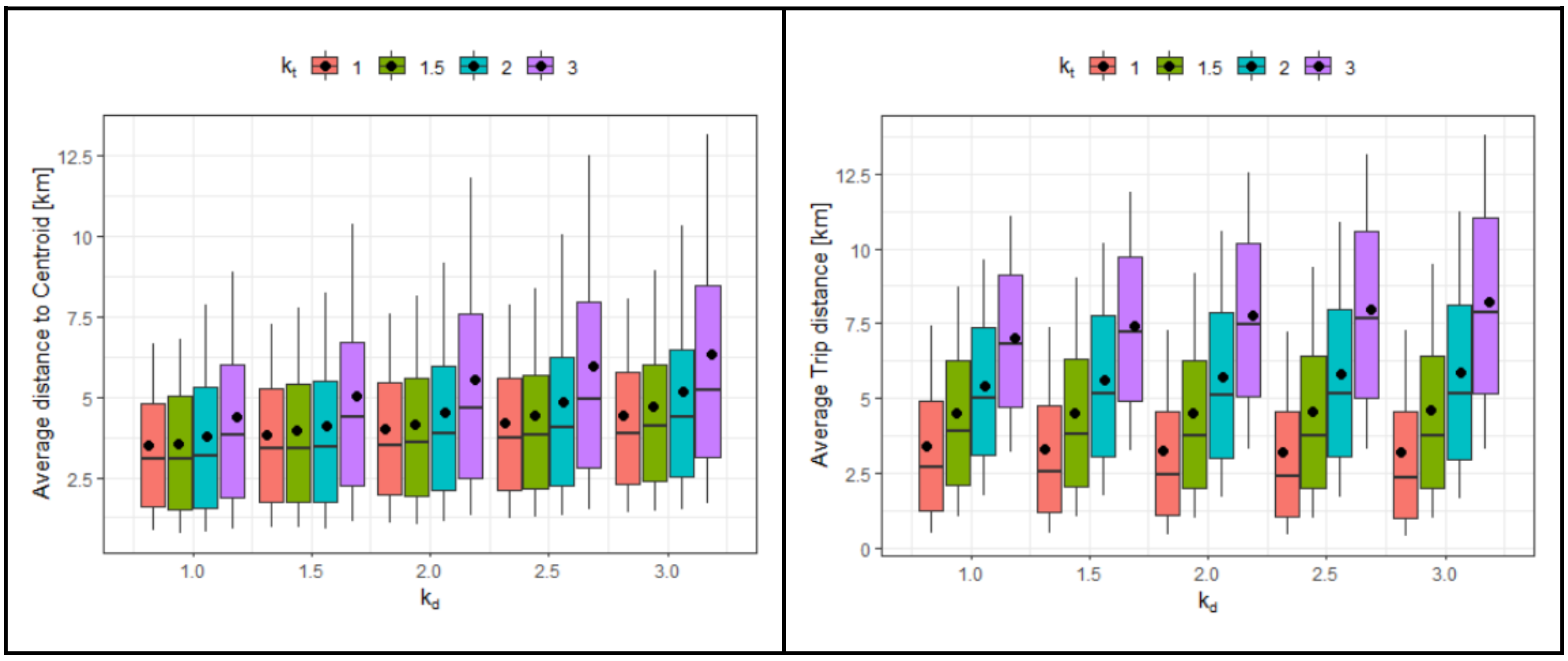

**Figure 4 Distance between generated destinations and a single center (Dam Square) (left), and trip length distribution (right)**

In addition, we display heatmaps of the geographical distribution of the generated demands. In Figure 5, the four different $k_d$ values are presented for a fixed $k_t$. As expected, we see how increasing this parameter increases the spread of the generated destinations across the city. We carry out the same analysis for the four different $k_t$ values and a fixed $k_d$, as presented in Figure 6. We observe a similar trend as for the previous heatmaps, yet less pronounced. We also see a larger number of trips heading to the peripheral areas of the city, due to the increase in trip length distribution parameter. Finally, we conduct a similar analysis for a specific combination of $k_t$ and $k_d$ (both equal to 1.5) but for different numbers of centers of attraction, presented in Figure 7. We observe, as expected, how increasing the number of centers increases the spread of the demand destinations generated, yet visibly concentrated around the respective centers (marked with purple triangles).

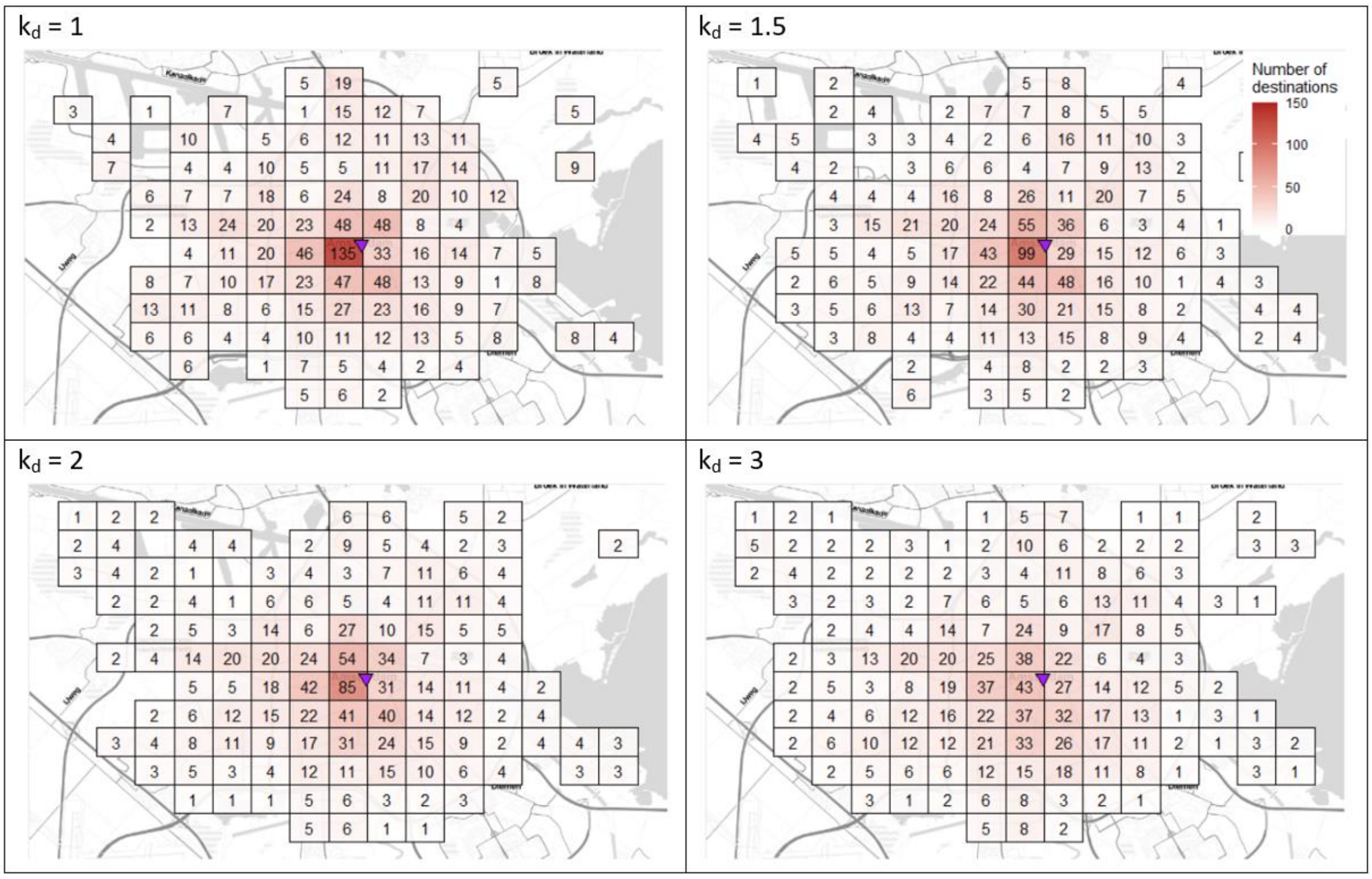

**Figure 5 Synthetic demand patterns. Spatial distribution of destinations of various concentrations around centers ($k_d$). One center (Dam Square) and $k_t$ = 1.5.**

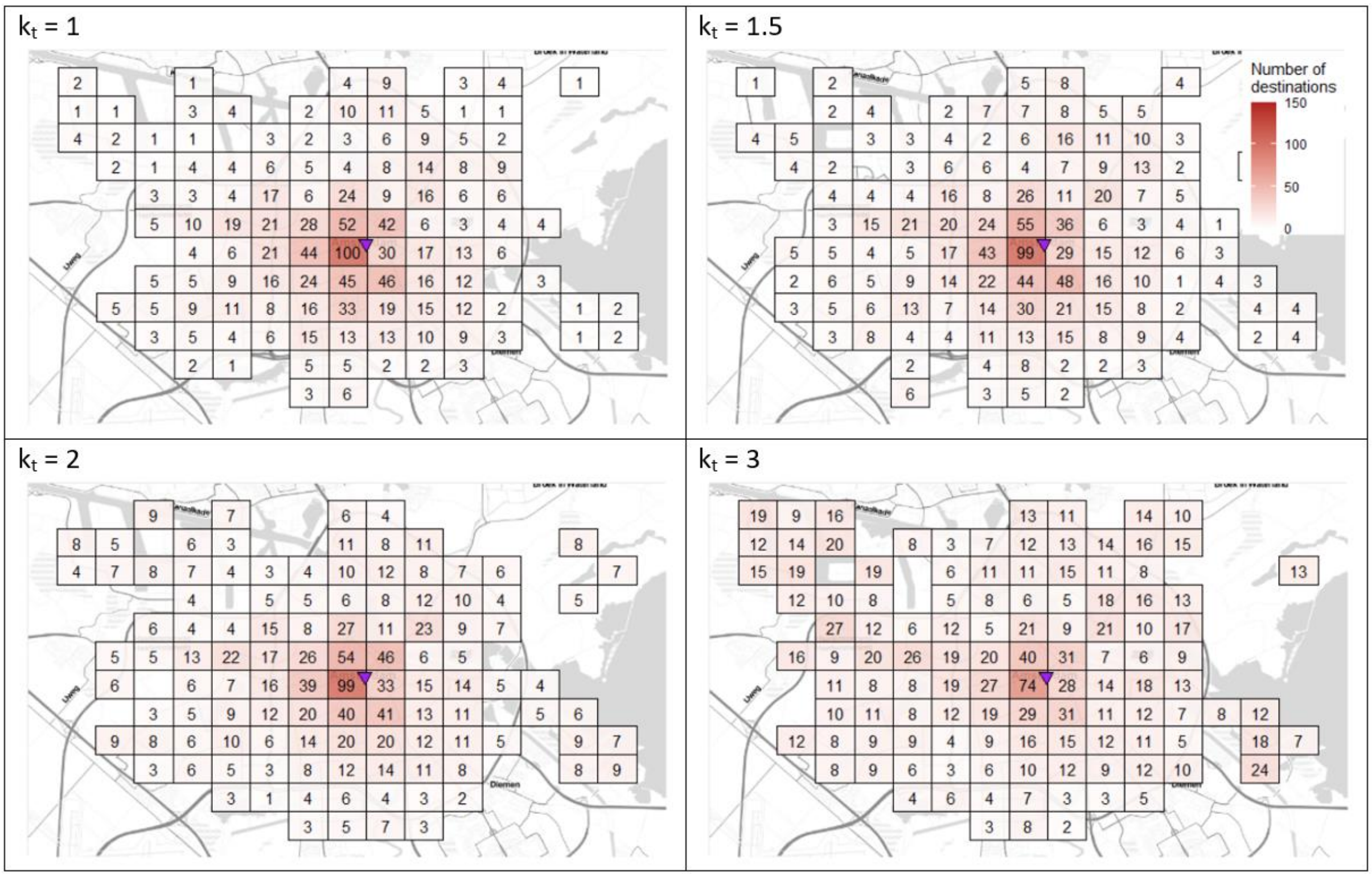

**Figure 6 Synthetic demand patterns. Spatial distribution of destinations of various trip length distribution ($k_t$). One center (Dam Square) and $k_d$ = 1.5.**

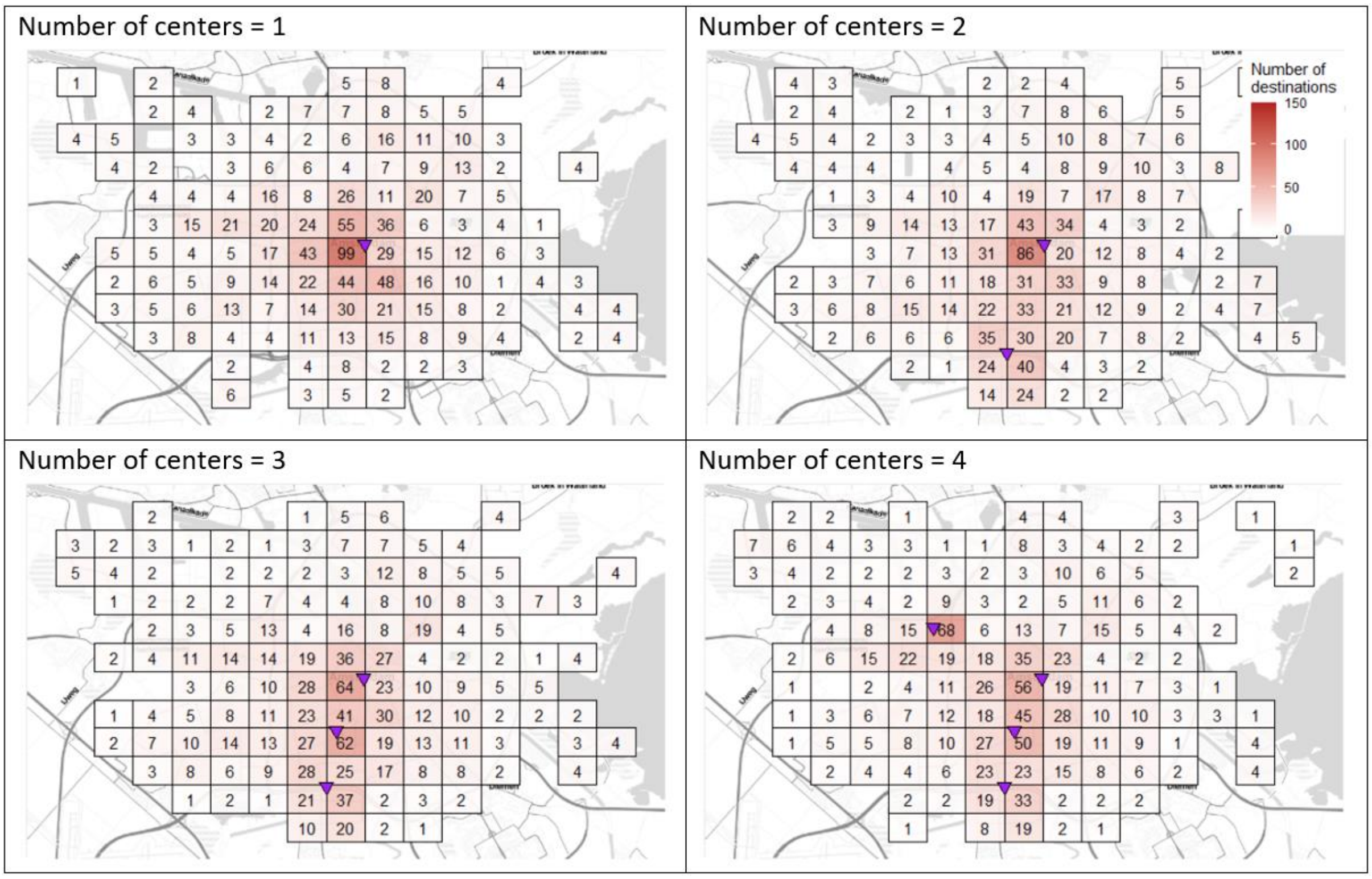

**Figure 7 Synthetic demand patterns. Spatial distribution of destinations for different number of centers ($k_t$ = 1.5 and $k_d$ = 1.5)**

### 3.2.2 Aggregate shareability results

The demand pattern generation scenarios not only yield substantially different trip characteristics (as reported in the previous subsection), but also lead to significantly different shareability prospects as we report in Table 1. Recall that all of the results presented are averaged over 10 scenario replications, each of which involves sampling from the respective demand distribution settings. The occupancy varies from 1.25 in the least shareable scenario to 1.75 in the most shareable one. Vehicle hours can be reduced by 18% in the least shareable scenario and up to almost 60% in the most shareable scenarios. The impacts of introducing ride-pooling on passenger costs varies between a reduction of 2.1% and 2.9%.

Notably, not all parameters of demand pattern distribution have the same impact. While destinations' density $k_d$ and the number of centers have a relatively limited impact, we find that all the indicators are most sensitive to $k_t$. For a monocentric demand with $k_d = 1$ the shareability increases with $k_t$ as reflected across all performance metrics. The vehicle hours reduction more than doubles when $k_t$ increases from 1 to 3 and thus trip lengths increase. Mind that vehicle hours in private (non-pooled) scenario also doubles when $k_t$ increases from 1 to 3. We find that $k_d$ and the number of centers are negatively associated with travel time and thus with shareability. Changing $k_d$ from 1 to 3 changes the total travel time from 118.3 hours to 111.1 hours, which corresponds to the travel time reduction attributed to sharing dropping from 23% to 19%. Similarly, passenger cost reduction diminishes from 2.27% to 2.09%. Shareability metrics are overall stable in relation to changes in the number of centers where the total travel time reductions not falling below 20% and passenger costs reductions remain greater than 2.15% even for scenarios with four centers.

| | N° centers | $k_d$ | $k_t$ | O | Vehicle hours *T* | | | Passenger costs *C* | | |
|---|---|---|---|---|---|---|---|---|---|---|
| | | | | | private | pooled | rel diff. Δ | private | pooled | rel diff. Δ |
| **Modifying density $k_t$** | 1 | 1 | 1 | 1.30 | 426.573 | 346.641 | -23% | 6.617 | 6.470 | -2,27% |
| | 1 | 1 | 1.5 | 1.42 | 559.537 | 420.776 | -33% | 8.678 | 8.460 | -2,58% |
| | 1 | 1 | 2 | 1.54 | 676.868 | 476.158 | -42% | 10.497 | 10.217 | -2,74% |
| | 1 | 1 | 3 | 1.74 | 878.131 | 551.206 | -59% | 13.616 | 13.237 | -2,86% |
| **Modifying trip distance $k_d$** | 1 | 1 | 1 | 1.30 | 426.573 | 346.641 | -23% | 6.617 | 6.470 | -2,27% |
| | 1 | 1.5 | 1 | 1.29 | 414.04 | 340.386 | -22% | 6.423 | 6.282 | -2,24% |
| | 1 | 2 | 1 | 1.26 | 405.519 | 339.467 | -19% | 6.291 | 6.158 | -2,16% |
| | 1 | 3 | 1 | 1.25 | 400.748 | 339.51 | -18% | 6.217 | 6.090 | -2,09% |
| **Modifying the number of centers** | 1 | 1 | 1 | 1.30 | 426.573 | 346.641 | -23% | 6.617 | 6.470 | -2,27% |
| | 2 | 1 | 1 | 1.29 | 422.477 | 345.215 | -22% | 6.554 | 6.406 | -2,31% |
| | 3 | 1 | 1 | 1.28 | 406.773 | 335.664 | -21% | 6.310 | 6.169 | -2,29% |
| | 4 | 1 | 1 | 1.28 | 400.163 | 329.123 | -22% | 6.208 | 6.076 | -2,17% |

**Table 1. Shareability indicators for various demand patterns.**

We further analyse the underlying relations between demand generation parameters and the shareability indicators by plotting related trends. In Figure 8 we show how the three shareability metrics – *O*, *T* and *C* - change under various demand configurations for a monocentric city. While for lower values of impedance ($k_t$<3) increasing dispersion from the centers has a negative impact for all metrics, for $k_t$=3 the trend is inverted for occupancy and vehicle hours reduction. For instance changing $k_d$ from 1 to 3 barely changes both the occupancy and vehicle hours reduction, whereas changing $k_t$ from 1 to 3 changes it substantially: occupancy rises from 1.25 to 1.8 and the reduction in vehicle hours increases from 15% to almost 40%. The trends are consistent with the exception of passenger costs (right panel), where for high values of $k_t$ (2 and 3), the positive impact of sharing on passenger cost first increases and then decreases. The spacings between the lines (values of $k_t$) are similar for occupancy and vehicle hours reduction, with $k_t$=3 being a clear outlier. Furthermore, for passenger costs $k_t$=1 is the outlier with the least positive impact (differences between $k_t$=2 and $k_t$=3 are smaller than between $k_t$=1 and $k_t$=1.5).

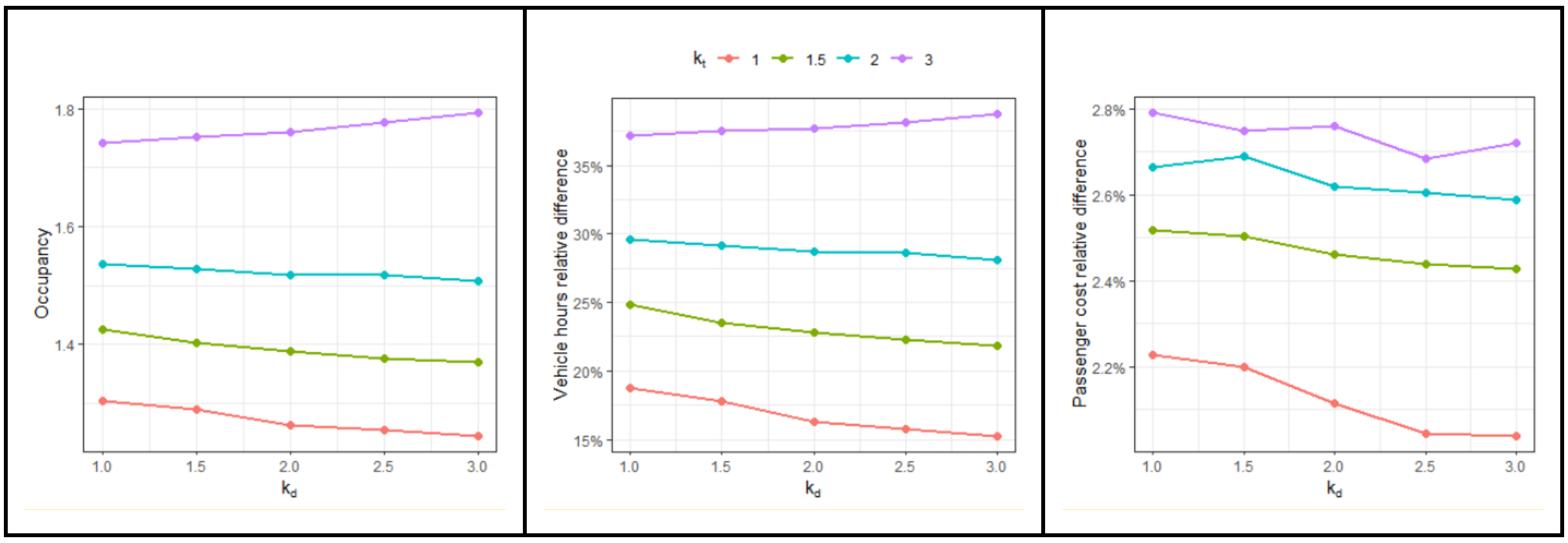

**Figure 8 Occupancy (left), reduction in vehicle hours (middle) and reduction in passenger costs (right) in a monocentric city with varying distance from centers and trip length distribution**

Results for more polycentric cities may significantly vary from those for monocentric ones. We report here the changes in the shareability metrics for different numbers of centers, first for shorter ($k_t$=1) and longer trips ($k_t$=3) in Figure 9. When the trips are short, results are less sensitive to $k_d$, the more centers are present. While for a monocentric city vehicle hour reduction varies with $k_d$ from 15% to 19.5%, for three and more centers the difference narrows by almost twice, from 15.5% to 17.5%. Nonetheless, in the case of short trips, a monocentric city typically provides the best results if $k_d<2$ and introducing new centers limits the benefits of sharing. In contrast, for $k_d \geq 2$ trends are not that clear. For instance with $k_d=2.5$ introducing a second center increases passenger benefits, while the introduction of a third or a fourth one reduces the passenger benefits attained. Conversely, the reduction in vehicle hours declines after introducing the second center when $k_d=2.5$ to then improve when a third or a fourth center are introduced.

When travellers perform longer trips, the benefits stemming from sharing are in general greater (see also Table 1). However, the changes with respect to the number of centers are somehow different. Cities with four centers yield the greatest reductions in vehicle hours and the highest occupancy, while duo-centric cities yield the lowest benefits. Introducing a second center worsens both occupancy and vehicle hours reductions for all cases except for when $k_d = 1$. For cities with four centers and $k_d \geq 2$ passenger benefits are lowest and drop significantly when the number of centers changes from 2 to 4, while introducing a second center increases passenger benefits. Unlike for $k_t = 1$, where trends across the three metrics are consistent, under $k_t$=3 contradicting trends among those can be observed. For instance, when $k_d = 2$ both occupancy and vehicle hour reductions are greatest for cities with two centers, while for passenger costs it is the opposite and the worst results are yielded when there are two prime centers.

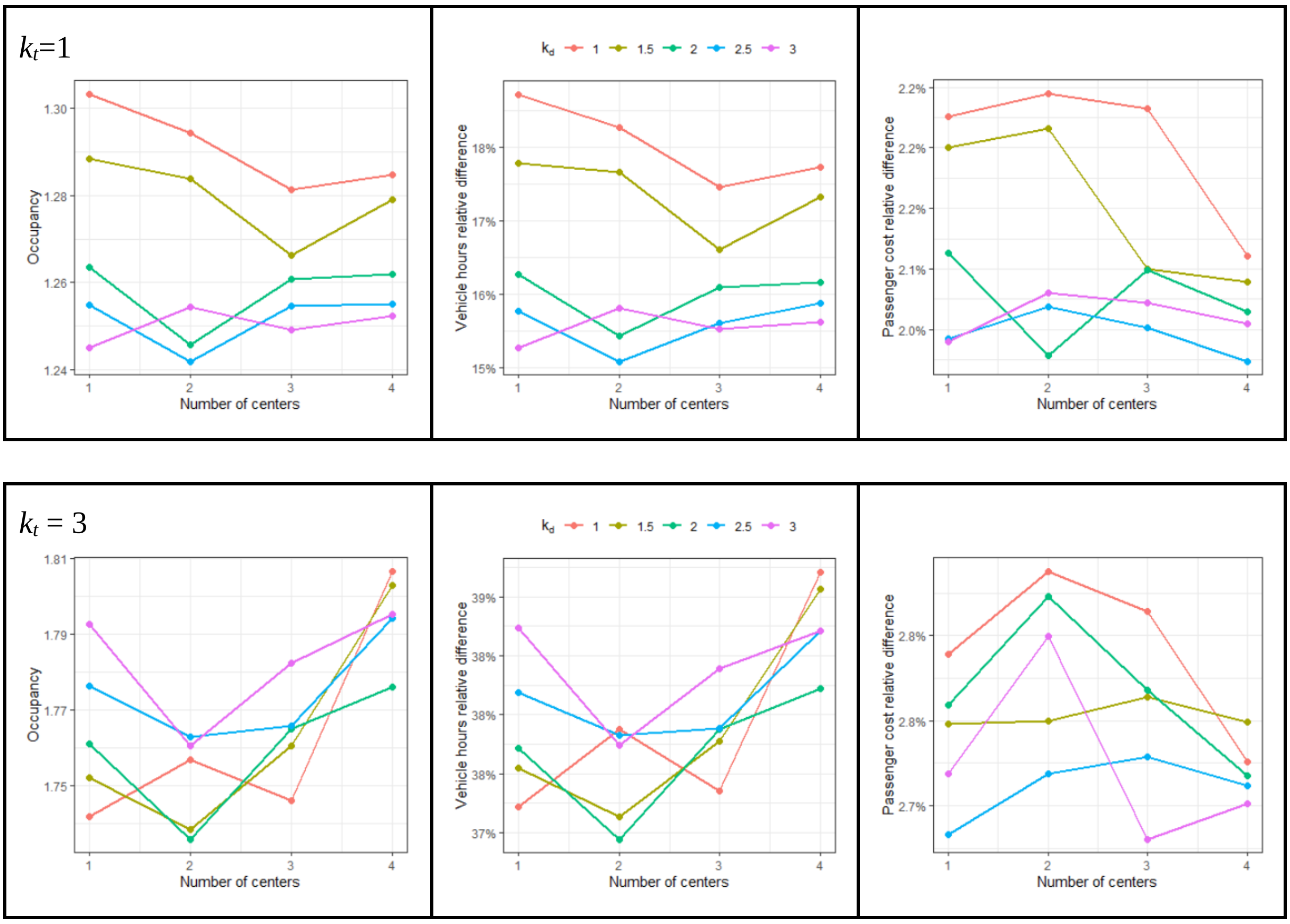

**Figure 9 Occupancy (left), reduction in vehicle hours (middle) and reduction in passenger costs (right) for urban areas with a different number of centers and destination density under $k_t$ = 1 (above) and $k_t$ = 3 (below).**

### 3.2.3 Distributional effects of shareability

In the previous subsection we have analysed the overall relations between demand pattern parameters and the shareability metrics. Next, we investigate how the shareability potential ingrained in each of the demand scenarios varies among users and manifests itself spatially as well as the extent to which spatial disparities are exhibited. The analysis hence sheds light on the distributional effects associated with the introduction of ride-pooling. In the following, we present and discuss the histogram of each of the performance indicators as well as display average values calculated over trip origins for each grid cell by overlaying a grid of 1.33 squared-km over the case study area. We start by examining the impact that $k_t$ has in a monocentric city, i.e. how much it affects longer travel distances, and then follow with an analysis of the impact of having a certain number of centers of attraction.

The distributions of the shareability metrics are presented in Figure 10. Instead of studying the occupancy as in the previous sections, we analyse the share of trips that have been successfully matched with at least one co-traveller. This is done as it not possible to account for the passenger and vehicle times for each trip request (i.e. the nominator and denominator of occupancy, *O*, respectively) in a consistent way. The histograms present the distributions of each metric across the grid cells, varying from 0 if none of trips originating from the cell were shared up to 100% where each traveller finds a match. As discussed in the previous subsection, we see how there is a clear increase in the extent to which people share their rides when increasing the travelled distance. The longer the distance, the higher the likelihood that (parts of) trips can be shared. In addition, we see that there is also a higher variability, both at a ride level and at the geographical grid cell level when $k_t$ is lower. In this scenario, we observe that a higher travel distance distribution leads also to a more homogeneous (and higher) shareability levels across the city and across users.

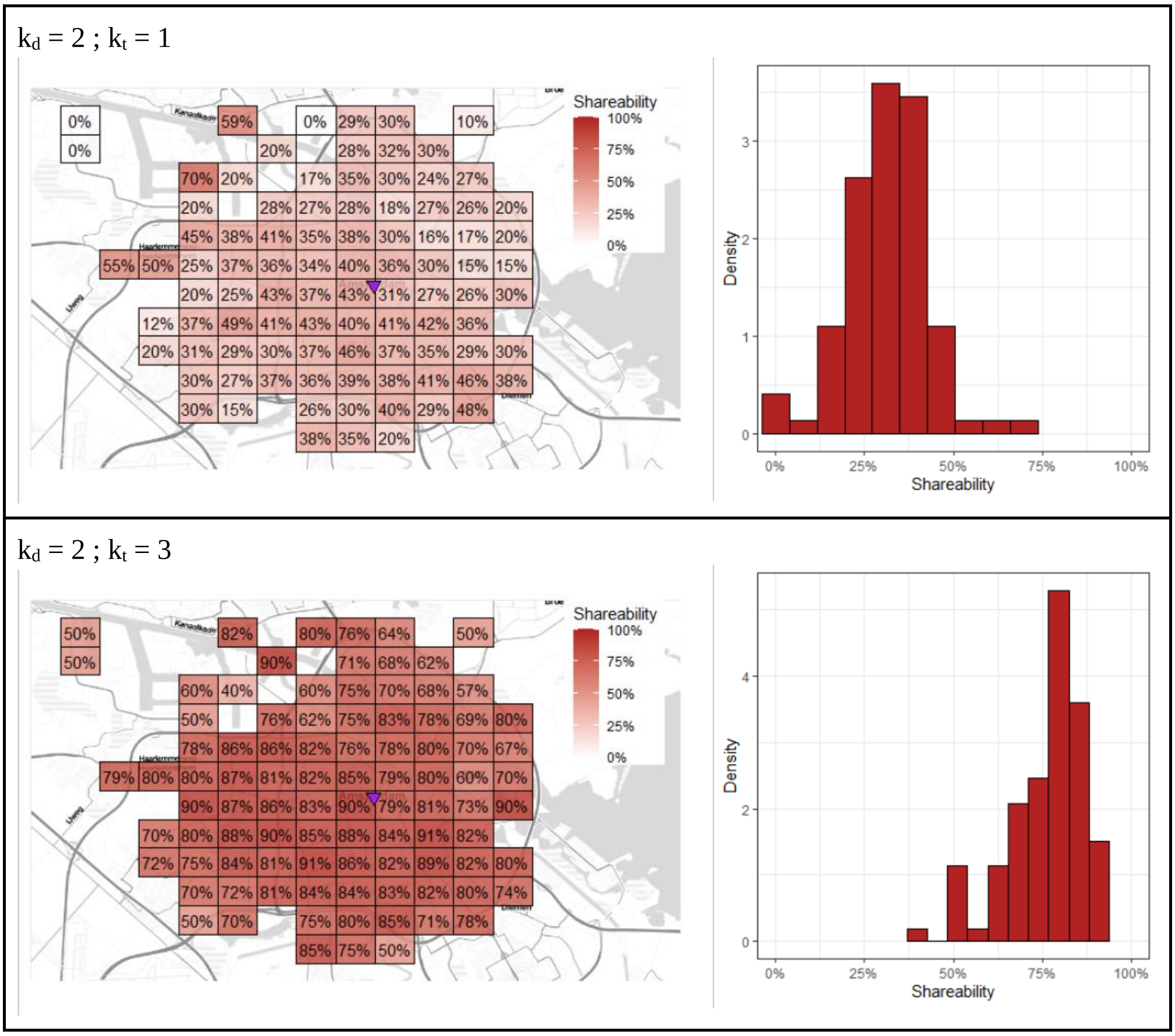

**Figure 10 Distributional analysis of the percentage of trips shared in a monocentric city under two travel distances distribution scenarios: with short trips (top) and long trips (bottom)**

When considering the differences in total vehicle hours, we observe a similar trend in the distribution. This is expected, as there is a direct relation between travel distance and resulting vehicle-hours. However, in this case, we see a higher variability for the case of higher $k_t$. As more people are willing to travel to the center of attraction, and also travel distances are longer, the extent of the detour becomes more variable. Mind that passenger experience no detour when either starting close enough to their destination in a shared ride or when they did not find a match and travelled private.

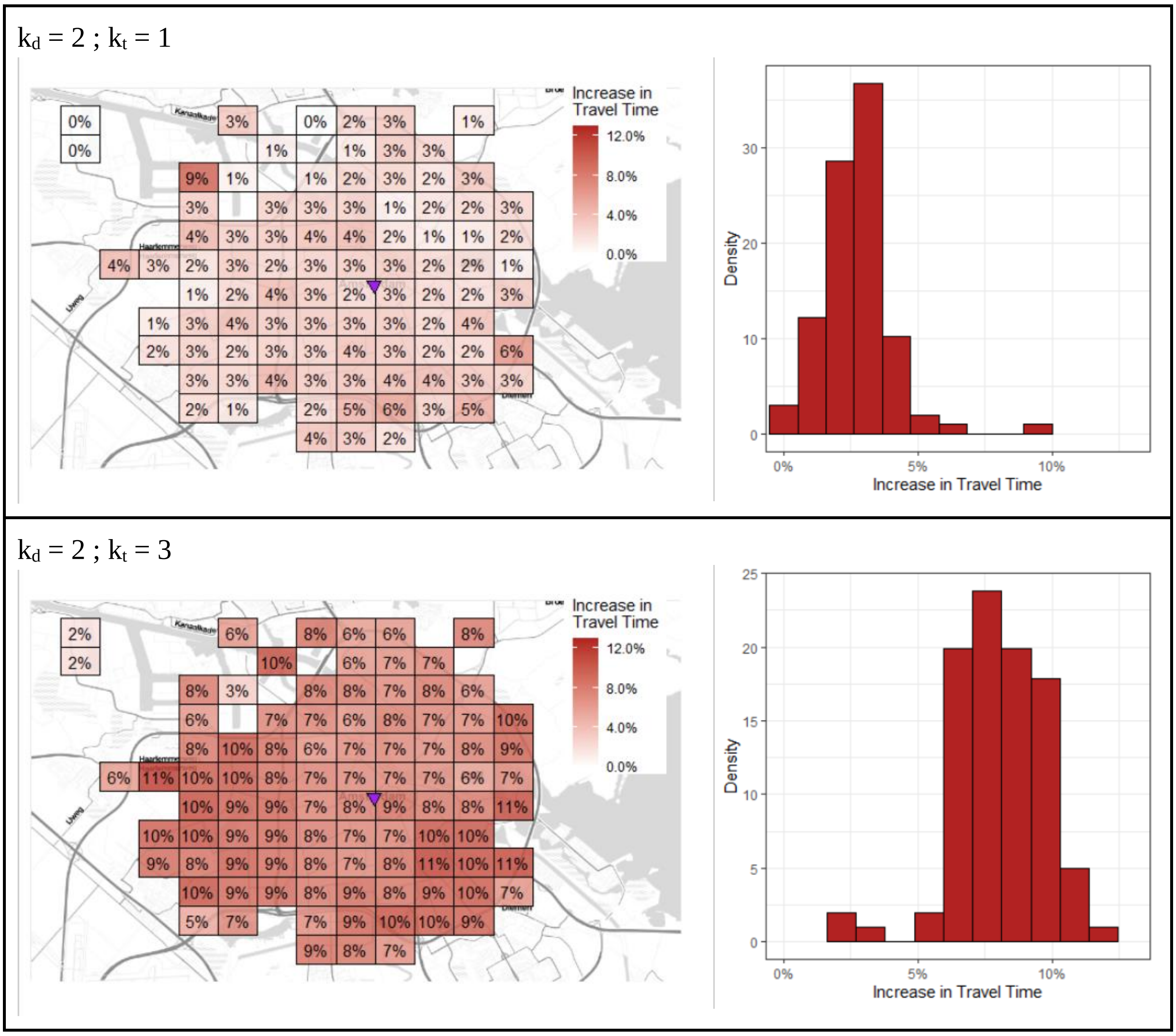

**Figure 11 Distributional analysis of the increase in vehicle hours following the introduction of ride-pooling in a monocentric city under two travel distances distribution scenarios**

In order to consider the two already mentioned effects together, we analyse the distributional effect of the total passenger costs. Overall, we observe a similar trend as above, meaning an increase in the shareability indicator value when $k_t = 3$. However, the differences in this scenario are considerably smaller because of the discount benefits being compensated by the prolonged travel times when more people are inclined to travel longer distances. Variability in this case is less pronounced and the differences between the two cases are modest.

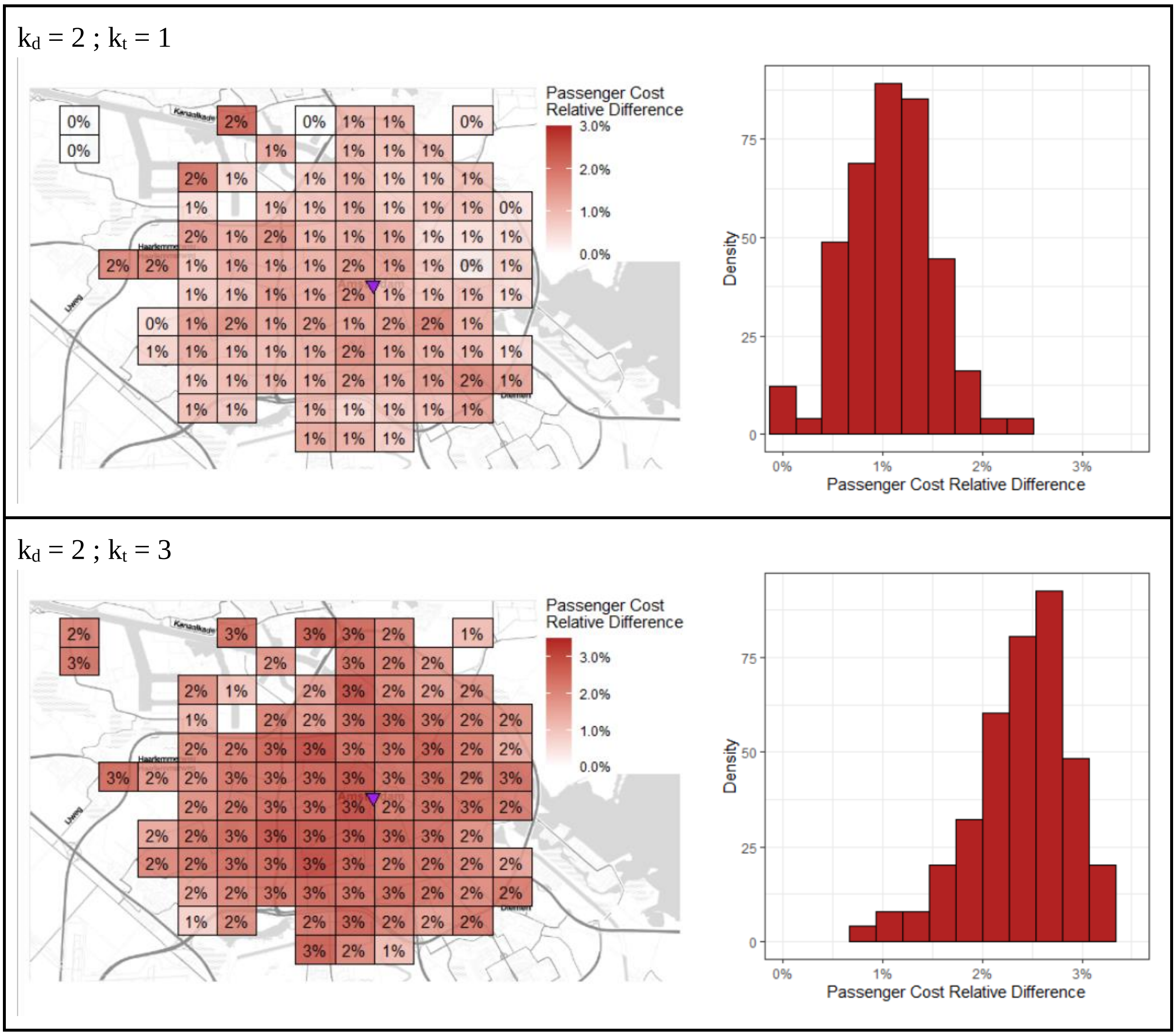

**Figure 12 Distributional analysis of the reduction in passenger costs following the introduction of ride-pooling in a monocentric city under two travel distances distribution scenarios**

Finally, we consider the scenario of a simulated polycentric city with four centers of attractions and its effects on shareability. When compared to the heatmap and histogram shown in Figure 10, we see that share of shared trips is lower in this scenario then in the corresponding monocentric scenario. This is expected as destinations are concentrated in four different areas instead of only in one, which reduces the chances for people to have mutually compatible rides. In addition, we see that there is a higher heterogeneity across the city, which is related to the proximity of each origin to the potential destinations. People in the peripheral areas now have a reduced chance to share their rides because both the chance of finding a feasible sharing is lower and also the detour penalty is higher for those, resulting in less attractive matches and hence the ride-pooling alterative is less likely to be selected over private rides.

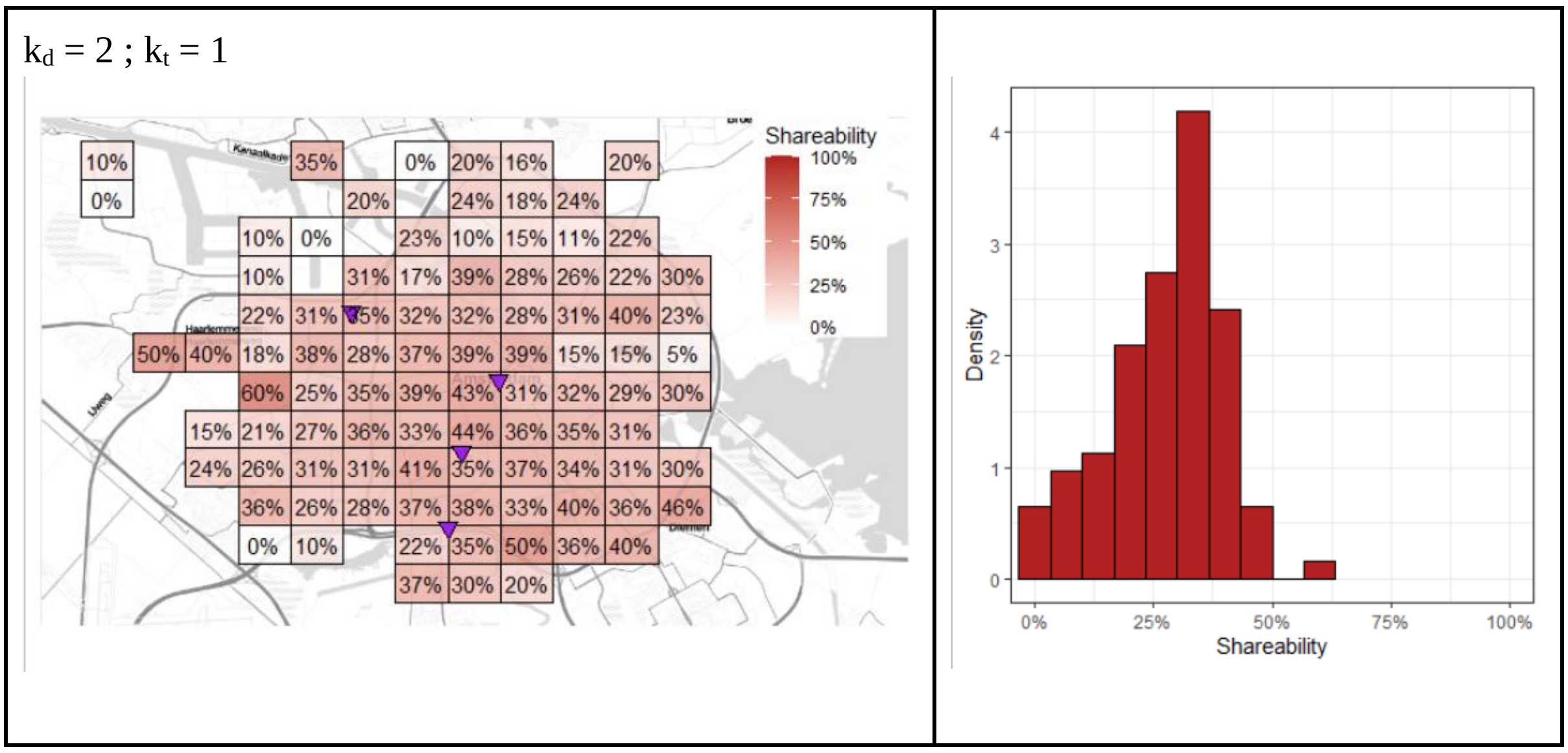

**Figure 13 Distributional analysis of the percentage of trips shared in a polycentric urban area with four centers under two travel distances distribution scenarios**

## 4. Discussion and conclusions

Shared rides are often considered to be a promising travel alternative that could efficiently pool people together while offering a door-to-door service. Despite this premise, when examining large-scale ride-pooling services in large urban areas, they are either not yet available or have not claimed a significant market share yet. For ride-hailing platforms to offer a ride-pooling alternative, it is of utmost importance to ensure that demand levels are sufficiently high so that there is a non-negligible probability that trips could be matched. Otherwise, the platform bears the risk of offering considerable discounts to passengers that end up travelling on their own (even though they signed up for a shared ride service) without saving on driver commissions. The likelihood that trip requests can be matched depends on their mutual compatibility in terms of trips' origin, destination and departure time. While this mutual compatibility is an essential prerequisite, it is by no means sufficient. Hence, when matching trips into rides in our study, we do not only ensure their mutual compatibility in time and space. Instead, we also guarantee that shared rides are only composed by travellers that find the ride-pooling offer to be more attractive than the private ride-hailing alternative given the trade-offs between travel time, fare and discomfort.

Our analysis sheds light on the consequences of the spatial demand pattern on the resulting prospects for a ride-pooling service. On one hand, a demand pattern which is more concentrated in space is likely to allow for more opportunities for matching similar trips without inducing substantial detours. On the other hand, the incentive to share a ride is greater for longer rides due to the discount offered. There is hence a non-trivial relation between demand pattern characteristics and its potential shreability which calls for a detailed empirical assessment. Our findings indicate that introducing a ride-pooling service can reduce vehicle-hours by 18-59 %, depending on the concentration of travel destinations around the center and the trip length distribution. System efficiency correlates positively with the former and negatively with the latter. The respective on-board occupancy level varies from 1.25 to 1.74. Passenger costs savings are hovering between 2-3 % for all scenarios. These gains are attained by passengers who shift from private to shared rides because they find that the discount more than compensates for the detour and discomfort induced (otherwise they will not have opted for the ride-pooling alternative).

Interestingly, the number of major centers of attraction across the city does not have a considerable difference, everything else being equal.

The results suggest that cities characterised by compact center(s) of attraction and where travellers are inclined to use ride-hailing for those trips where destinations are located further away from their origins, i.e. in our case study of Amsterdam this corresponds to average trip distance of 5-8 km, offer the most fertile ground for attracting users to share their rides and result with an increased efficiency in on-demand transport services. Our analysis of the distributional effects in shareability levels under different demand patterns also reveal that those patterns that induce most ride-pooling are also those that result with the most even distribution of service performance - both across the population as well as across different parts of the urban area. This demonstrates that the impacts of ride-pooling are not limited to selected localized effects but are rather well distributed because the efficiency gains are made possible by matching individuals with a diverse set of origins and destinations (in contrast to a shuttle service).

Further research may extend our analysis by sampling travel origins and examining the impacts of alternative trip generation and attraction patterns. This will allow for analyzing more complex spatial relations that extend beyond the (mono or poly)centric patterns assumed in this study. Similarly, the daily temporal variations in the demand patterns analysed in this study have been specified based on the travel demand profile data for Amsterdam. Future research may also examine how different temporal and spatial profiles, based on travel patterns observed for other cities, may impact the potential of ride-pooling services.

The approach adopted in this study can be coupled with models of innovation diffusion and supply evolution in order to identify the conditions relevant for obtaining a critical mass of both users and drivers in a two-sided on-demand transport platform. Moreover, embedding our method in a travel demand model that includes a feedback loop to modal choices will allow treating the total demand for on-demand transport services as an endogenous variable and hence conclude on the ability of ride-pooling services to attract users which will in its absence use alternative means of travel, other than private ride-hailing.

## Acknowledgements

This research was supported by the CriticalMaaS project (grant number 804469), which is financed by the European Research Council and Amsterdam Institute for Advanced Metropolitan Solutions.

**References**

Arentze, T. A. and Timmermans, H. J. (2004). A learning-based transportation oriented simulation system. *Transportation Research Part B*, 38(7), 613-633.

Badia H. (2020). Comparison of bus network structures in face of urban dispersion for a ring-radial city. *Networks and Spatial Economics*, 20, 233-271.

Bassolas A., Barbosa-Filho H., Dickinson B., Dotiwalla X., Gallotti R., Ghoshal G., Gipson B., Hazarie S.A., Kautz H., Kucuktunc O., Lieber A., Sadilek A. and Ramasco J.J. (2019). Hierarchical organization of urban mobility and its connection with city livability. *Nature Communications*, 10, 4817.

Blafoss Ingvardson J. and Anker Nielsen O. (2018). How urban density, network topology and socio-economy influence public transport ridership: Empirical evidence from 48 European metropolitan areas. *Journal of Transport Geography*, 72, 50-63.

Boisjoly G., Grisé E. Maguire M., Veillette M-P. Deboosere R., Berrebi E. and A. El-Geneidy (2018). Invest in the ride: A 14 year longitudinal analysis of the determinants of public transport ridership in 25 North American cities. *Transportation Research Part A*, 116, 434-445.

Cats O., Vermeulen A., Warnier M. and van Lint H. (2020). Modelling Growth Principles of Metropolitan Public Transport Networks. *Journal of Transport Geography*, 82, 1-10.

Erhardt G.D., Sneha R., Copper D., Sana B., Chen M. and Castliglione J. (2019). Do transportation network companies decrease or increase congestion? *Science Advances*, 5 (5), eaau2670.

Ewing R. and Cervero R. (2010). Travel and the built environment, a meta-analysis. *Journal of the American Planning Association*, 76 (3), 265-294.

Ewing R., Tian G. and Lyons T. (2018). Does compact development increase or reduce traffic congestion? *Cities*, 72 (A), 94-101.

Fielbaum A., Jara-Diaz S. and Gschwender A. (2016). Optimal public transport networks for a general urban structure. *Transportation Research Part B*, 94, 298-313.

Li W., Pu Z., Li Y. and X.J. Ban (2019a). Characterization of ridesplitting based on observed data: a case study of Chengdu, China. *Transportation Research Part C*, 100, 330-353.

Li Y., Xiong W. and Wang X. (2019b). Does polycentric and compact development alleviate urban traffic congestion? A case study of 98 Chinese Cities. *Cities*, 88, 100-111.

Narayan, J., Cats, O., van Oort, N., & Hoogendoorn, S. (2021). On the scalability of private and pooled on-demand services for urban mobility in Amsterdam. Under review.

Oke J.B., Aboutaleb Y.M., Akkinepally A., Lima Azevedo C., Han Y., Zegras P.C., Ferreira J. and Ben-Akiva M.E. (2019). A novel global urban topology framework for sustainable mobility futures. *Environment Research Letters*, 14, 095006.

Tachet R., Sagarra O., Santi P., Resta G., Szell M., Strogatz S. H. and Ratti C. (2017). Scaling law of urban ride sharing. *Scientific Reports*, 7, 42868.

Tsekeris T. and Geroliminis N. (2013). City size, network structure and traffic congestion. *Journal of Urban Economics*, 76, 1-14.

Wang K. and Zhang W. (2020). The role of urban form in the performance of shared automated vehicles. https://arxiv.org/pdf/2012.01384.pdf

Young M., Farber S. and Palm M. (2020). The true cost of sharing: A detour penalty analysis between UberPool and UberX trips in Toronto. *Transportation Research Part D*, 87, 102540.